\documentclass[12pt]{article}

\usepackage{graphicx}
\usepackage{amsmath, amssymb}
\usepackage{slashed}
\usepackage{color}
\usepackage{bm}
\usepackage{cite}
\usepackage{comment}
  
\allowdisplaybreaks[1]

\begin{document}

\begin{titlepage}

\def\thefootnote{\fnsymbol{footnote}}

\begin{center}

\hfill TU-998\\
\hfill KANAZAWA-15-10 \\
\hfill September 4, 2015

\vspace{0.5cm}

{\Large\bf Direct detection of singlet dark matter \\ in  classically
  scale-invariant standard model}

\vspace{1cm}
{\large Kazuhiro Endo}$^{\it (a)}$, 
{\large Koji Ishiwata}$^{\it (b)}$

\vspace{1cm}

{\it $^{(a)}${Department of Physics, Tohoku University, Sendai
    980-8578, Japan}}

\vspace{0.2cm}

{\it $^{(b)}${Institute for Theoretical Physics, Kanazawa University,
    Kanazawa 920-1192, Japan}}

\vspace{0.2cm}

\vspace{1cm}

\abstract{ Classical scale invariance is one of the possible solutions
  to explain the origin of the electroweak scale. The simplest
  extension is the classically scale-invariant standard model
  augmented by a multiplet of gauge singlet real scalar. In the
  previous study it was shown that the properties of the Higgs
  potential deviate substantially, which can be observed in the
  International Linear Collider. On the other hand, since the
  multiplet does not acquire vacuum expectation value, the singlet
  components are stable and can be dark matter. In this letter we
  study the detectability of the real singlet scalar bosons in the
  experiment of the direct detection of dark matter. It is shown that
  a part of this model has already been excluded and the rest of the
  parameter space is within the reach of the future experiment.  }

\end{center}
\end{titlepage}


\renewcommand{\theequation}{\thesection.\arabic{equation}}
\renewcommand{\thepage}{\arabic{page}}
\setcounter{page}{1}
\renewcommand{\thefootnote}{\#\arabic{footnote}}
\setcounter{footnote}{0}

\section{Introduction}
\label{sec:intro}
\setcounter{equation}{0} 

Higgs boson was discovered in 2012 at the CERN Large Hadron
Collider~\cite{Aad:2012tfa, Chatrchyan:2012ufa}. Since then, its
properties, such as spin, parity and couplings to the standard model
fermions and gauge bosons, have been measured and it turned out that
they are consistent with the standard model prediction. In spite of
the success of the standard model up to now, however, it is commonly
believed that the standard model is not the ultimate theory of
particle physics. In fact there are lots of unsolved problems in the
field of particle physics as well as cosmology.

One of them is the origin of the spontaneous symmetry breakdown of the
electroweak gauge group. In the standard model, the electroweak
symmetry is broken by Higgs field that has an ad hoc tachyonic mass
term. One explanation for the tachyonic mass is supersymmetry.  In
supersymmetric extension of the standard model, the negative mass term
is induced radiatively. On the other hand, radiative symmetry breaking
is possible in non-supersymmetric theory, which is known as
Coleman-Weinberg (CW) mechanism~\cite{Coleman:1973jx}. In the CW
mechanism spontaneous symmetry breaking is induced at quantum level
from classically scale-invariant scalar potential. Although it turned
out that the CW mechanism with a Higgs does not work for the
electroweak symmetry breaking, simple extensions of the Higgs sector
are known to be phenomenologically viable (see, {\it e.g.},
\cite{Foot:2007as,Espinosa:2007qk,AlexanderNunneley:2010nw,Dermisek:2013pta,Masina:2013wja,Antipin:2013exa,Guo:2014bha,Hempfling:1996ht,Tamarit:2014dua,Chang:2007ki,Meissner:2006zh,Foot:2007iy,Iso:2009ss,Foot:2010av,Ishiwata:2011aa,Iso:2012jn,Englert:2013gz,Hambye:2013dgv,Carone:2013wla,Gabrielli:2013hma,Abel:2013mya,Radovcic:2014rea,Khoze:2014xha,Farzinnia:2014xia,Altmannshofer:2014vra,Benic:2014aga}).

Recently Higgs properties were studied in a classically scale-invariant 
standard model augmented by an electroweak singlet scalars that form a multiplet of global $O(N)$ symmetry~\cite{Endo:2015ifa}. It was shown that the Higgs 
self-couplings deviate significantly from the standard model
prediction. Such feature can be observed as a prominent signal of this
model at the next-generation lepton collider experiment, 
such as the International Linear Collider (ILC) 
\cite{Baer:2013cma,Asner:2013psa,Fujii:2015jha}. 
On the other hand,
it was also shown that the singlet field does not get a vacuum
expectation value (VEV). Then, other Higgs properties are unaffected
since there is no mixing between the singlet and Higgs. Another
important consequence is the stability of the singlet field due
to unbroken $O(N)$ symmetry. If the reheating temperature of the
universe is higher than the mass of the singlet, the singlet field is
thermalized. Then non-vanishing thermal relic of the singlet remains,
which can play a role of dark matter.

In this letter we study direct detection of the real singlet dark
matter with $O(N)$ symmetry.  It was pointed out in a similar
framework where the thermal relic abundance of the singlet dark matter
is too small to explain the present energy density of dark matter by
taking into account the 125~GeV
Higgs~\cite{Dermisek:2013pta,Guo:2014bha,Tamarit:2014dua}. This is due
to an enhanced annihilation cross section caused by a large
singlet-Higgs coupling.  On the contrary, however, the large
singlet-Higgs coupling may result in a large scattering cross section
of the singlet with nucleon. According to the study of
Ref.\,\cite{Endo:2015ifa}, the couplings of the singlet with the
standard model particles are fixed for the successful electroweak
symmetry breaking via the CW mechanism, which makes it possible to
determine the relic abundance and the scattering cross section of the
singlet with nucleon at a high precision. The scattering cross section
of singlet scalar dark matter is also discussed in several literature
mentioned above. We revise the calculation of the spin-independent
cross section of singlet scalar particle by adopting the formalism
given in Ref.\,\cite{Hisano:2015rsa} where next-to-leading order QCD
effect is properly taken into account.  It will be shown that part of
the model has already been excluded by recent LUX
result~\cite{Akerib:2013tjd} and the future experiments will be able
to probe almost the entire parameter space of the model.

Here is the organization of this letter. In Sec.\,\ref{sec:model} we
briefly explain the model, including the prescription how to determine
model parameters. Then the thermal relic and the scattering cross
section of the singlet are calculated, and the detection of the
singlet scalar is discussed in
Sec.\,\ref{sec:DM}. Sec.\,\ref{sec:conclusion} is dedicated to
conclusion.

\section{The Model}
\label{sec:model}
\setcounter{equation}{0} 

In the framework with classical scale invariance it is known that the
standard model without the Higgs mass term has already been excluded.  In
order to construct phenomenologically viable model, therefore, it is
necessary to extend the model, {\it e.g.}, by adding a new particle to
the model.  The simplest extension is to introduce a gauge singlet
real scalar field.  Such a singlet scalar can couple to the Higgs in
general, then the singlet contributes to the CW potential.  The effect
strongly depends on the degree of freedom of the singlet field.  To
see the impact we introduce a fundamental representation of a global
$O(N)$ symmetry, $S=(S_1,\,\cdots,\,S_N)^T$.  Consequently, the
tree-level scalar potential which is allowed under the symmetry is
\begin{align}
V =
\lambda_H(H^\dagger H)^2
+\lambda_{HS}H^\dagger H\,S_iS_i
+\frac{\lambda_S}{4}(S_iS_i)^2\,,
\end{align}
where $H$ is the Higgs doublet field $H=(H^+,\,H^0)^T$, and summed
over $i=1,\,\cdots,\,N$ for $N\ge 2$. $Z_2$ symmetry is assumed for
$N=1$ case, whereas it is also a subgroup of $O(N)$ symmetry 
and always survives in the scale-invariant tree-level potential for $N\geq 2$.

The electroweak symmetry breaking is induced via the CW mechanism.  To
see this, the scalar fields can be taken without loss of generality as
$H=(1/\sqrt{2})\,(0,\,\phi)^T$ and $S =(\varphi,\,0,\,\cdots,\,0)^T$
where $\phi$ and $\varphi$ are classical fields of the real scalars.
Then the effective potential at one-loop level is given by
\begin{align}
V_\text{eff}(\phi,\,\varphi)
=V_\text{tree}(\phi,\,\varphi)
+V_\text{1-loop}(\phi,\,\varphi)\,,
\label{eq:effective_pot_unimp_complete}
\end{align}
with
\begin{align}
&V_\text{tree}(\phi,\,\varphi)
=\frac{\lambda_{H}}{4}{\phi}^4
+\frac{\lambda_{HS}}{2}{\phi}^2{\varphi}^2
+\frac{\lambda_{S}}{4}{\varphi}^4\,,
\label{eq:effective_pot_unimp_tree-level}
\\
&V_\text{1-loop}(\phi,\,\varphi)
=\frac{1}{4(4\pi)^2}\sum_{i}n_i{M_i}^4(\phi,\,\varphi)
\left[
  \ln \frac{{M_i}^2(\phi,\,\varphi)}{\mu^2}-c_i
  \right]\,,
\label{eq:effpot_ms-bar_mod}
\end{align}
in $\overline{\rm MS}$-scheme with renormalization scale $\mu$. Index
$i$ denotes the fields which run in the loop diagrams. ($n_i$,
${M_i}^2$ and $c_i$ are given in Appendix~\ref{sec:pot_param}.)  The
electroweak symmetry is spontaneously broken if $\partial V_{\rm eff}
/ \partial \phi|_{\phi=\langle \phi \rangle}=0$ with $\langle \phi
\rangle\neq 0$, which implies $\lambda_H\sim
\frac{N{\lambda_{HS}}^2}{16\pi^2}- \frac{3{y_t}^4}{16\pi^2}$ as a
necessary condition.  Then $\lambda_H$ should be regarded as the
next-to-leading order in terms of the order counting of the
dimensionless couplings. Consequently we rewrite the effective
potential as
\begin{align}
V_\text{eff}=V_\text{LO}+V_\text{NLO}\,,
\label{eq:0th+1st_perturbation_order}
\end{align}
with $V_\text{LO}$ and $V_\text{NLO}$ being regarded as leading order
(LO) and next-to-leading order (NLO) of the scalar potential;
\begin{align}
&V_\text{LO}=
\frac{\lambda_{HS}}{2}{\phi}^2{\varphi}^2
+\frac{\lambda_{S}}{4}{\varphi}^4\,,
\label{eq:0th_perturbation_order}
\\
&V_\text{NLO}=
\frac{\lambda_{H}}{4}{\phi}^4
\nonumber \\ &
+\frac{
  {F_{+\text{app}}}^{2}(\phi,\,\varphi)
}{64\pi^2}
\left[
  \ln\Bigl(\frac{F_{+\text{app}}(\phi,\,\varphi)}{\mu^{2}}\Bigr)
  -\frac{3}{2}
  \right]
+\frac{3}{64\pi^{2}}
\left({\lambda_{HS}}{\varphi}^2\right)^2
\left[\ln\Bigl(\frac{\lambda_{HS}{\varphi}^2}{\mu^2}\Bigr)
  -\frac{3}{2}
  \right]
\nonumber\\&
+\frac{N-1}{64\pi^{2}}
\left(\lambda_{HS}{\phi}^2+\lambda_{S}{\varphi}^2\right)^2
\left[\ln\Bigl(\frac{\lambda_{HS}{\phi}^2+\lambda_{S}{\varphi}^2}{\mu^2}\Bigr)
  -\frac{3}{2}
  \right]
\nonumber\\
&-\frac{12}{64\pi^2}{M_{t}}^{4}(\phi)
\left[\ln\Bigl(\frac{{M_{t}}^{2}(\phi)}{\mu^2}\Bigr)-\frac{3}{2}
  \right]
\nonumber\\&
+\frac{6}{64\pi^{2}}
    {M_W}^4(\phi)\left[\ln\Bigl(\frac{{M_W}^2(\phi)}{\mu^2}\Bigr)-\frac{5}{6}
      \right]
+\frac{3}{64\pi^{2}}{M_Z}^4(\phi)
\left[\ln\Bigl(\frac{{M_Z}^2(\phi)}{\mu^2}\Bigr)-\frac{5}{6}
	\right]\,,
\label{eq:1st_perturbation_order}
\end{align}
where $F_{+{\rm app}}$ is given in Appendix~\ref{sec:pot_param}. In
Ref.\,\cite{Endo:2015ifa} it is shown that the successful electroweak
symmetry breaking with the Higgs mass $m_h\simeq 125~{\rm GeV}$ can be
realized in a given number of $N$. Table~\ref{tab:parameter_01} shows
the results.\footnote{This is the results derived from the potential
  referred as (I) in Ref.\,\cite{Endo:2015ifa}.}  Roughly speaking,
the Higgs mass is expected to be $m_h\sim
(\sqrt{N}\lambda_{HS}/4\pi)v_H$ with $v_H=246~{\rm GeV}$, which is
consistent with the numerical results in
Table~\ref{tab:parameter_01}. With the proper order counting, the
effective potential around the VEV is obtained by replacing the scalar
fields as $\phi \to v_H +h$\,,~$\varphi^2 \to s_is_i$ and expanding by
powers of $h$ and $s_is_i$;
\begin{align}
V_\text{eff}
=&\, 
\text{const.}
+\frac{1}{2}{m_{h}}^2 h^2
+\frac{1}{2}{m_{s}}^2 \,s_is_i
+\frac{\lambda_{hhh}}{3!}\,v_{H}h^3
+\frac{\lambda_{hhhh}}{4!}h^4
\nonumber\\
&+\frac{\lambda_{hss}}{2}\,v_{H}\,h\,s_is_i
+\frac{\lambda_{hhss}}{4}h^2\,s_is_i
+\frac{\lambda_{ssss}}{4!}\,(s_is_i)^2
+\cdots ,
\label{eq:expandVeff}
\end{align}
where $m_s$ is the mass of singlet. We have taken $\mu=v_H$ and
omitted irrelevant terms in our later discussion. The results for
$\lambda_{hhhh}$, $\lambda_{hhh}$, $\lambda_{hhss}$, $\lambda_{hss}$,
$\lambda_{ssss}$ and $m_s$ are summarized in
Table~\ref{tab:parameter_02}~\cite{Endo:2015ifa}.\footnote{The
  couplings are obtained from the parameters shown in
  Table~\ref{tab:parameter_01} which corresponds to case (I) in
  Ref.\,\cite{Endo:2015ifa}.  Since the couplings change by a few \%
  in cases (II) or (III), we will use the couplings from case (I) in
  our later calculation.}  Basically $\lambda_{H}$ and $\lambda_{HS}$
are chosen to give rise to $v_H=246\,{\rm GeV}$ and $m_h=125\,{\rm
  GeV}$, which determine the couplings (except for the singlet
self-coupling) and the singlet mass.  $\lambda_{S}$, on the other
hand, has little impact on these results. Since the Higgs
self-couplings $\lambda_{hhhh}$ and $\lambda_{hhh}$ significantly
deviate from the SM prediction, the precise measurement of the Higgs
self-couplings is a viable way to test this model.

Another important fact shown in Ref.\,\cite{Endo:2015ifa} is that the
singlet does not get VEV.\footnote{ This fact is guaranteed to all
  orders in perturbative expansion.  Strictly speaking,
  non-perturbative effect might break $O(N)$, which could allow
  non-zero VEV for the singlet.  Possible (or known) non-perturbative
  effect is anomaly. In our model, however, $O(N)$ multiplet is
  scalar, thus it is anomaly free.  Though one may concern another
  unknown non-perturbative effect, the situation is the same for the
  standard model, {\it i.e.}  unbroken $U(1)_{\rm em}$ symmetry.}
Without a VEV of the singlet, Higgs properties, such as Higgs
production or decay rates, are unaffected.  On the other hand,
unbroken $O(N)$ symmetry forbids $s_i$ to decay. Such stable particles
can change the thermal history of the universe. If the reheating
temperature in the early universe is higher than the singlet mass, the
singlet particles are thermalized and their number densities freeze
out eventually. Then the thermal relics can be components of dark
matter.  In this model the parameters which determine the interaction
of $s_i$ with the standard model particles (the Higgs field in our
case) are completely fixed as discussed above. Therefore its nature is
highly predictable, and in fact we will see in the next section that
the experiments of direct detection of dark matter provide the
powerful tool to probe the model.

Before closing this section it is worth noting that $N=1$ case is
favored by two reasons \cite{Endo:2015ifa}.  In terms of Veltman's
condition~\cite{Veltman:1980mj} the level of fine-tuning for the Higgs
mass at the electroweak scale gets milder compared to larger $N$
cases. Second, the fine-tuning is relaxed compared to the standard
model and larger $N$ case in a sense that the cutoff scale due to
Landau pole is predicted to be around TeV.  We will come back to this
point later.

\begin{table}[t]
 \begin{center}
  \begin{tabular}{cccc}
   \hline\hline
   $N$ & 1 & 4 & 12 \\
   \hline\hline
   $\mu$ & \multicolumn{3}{c}{$v_H=246~\text{GeV}$} \\
   $y_t$  & \multicolumn{3}{c}{0.919} \\
   $g$ & \multicolumn{3}{c}{0.644} \\
   $g'$ & \multicolumn{3}{c}{0.359} \\
   $\lambda_{H}$  &$-0.11$ &$-0.0045$ &0.075 \\
   $\lambda_{HS}$  &4.8 &2.4 &1.4 \\
   $\lambda_{S}$  &0.10 &0.10 &0.10 \\
   \hline\hline
  \end{tabular}
  \caption{\small The input parameters of the analysis in
    Ref.~\cite{Endo:2015ifa}.  The values $\lambda_S=0.10$ are
    benchmarks, but there are very few $\lambda_S$-dependences for
    $\lambda_{hhh}$, $\lambda_{hhhh}$, $\lambda_{hss}$.}
 \label{tab:parameter_01}
 \end{center}
\end{table}

\begin{table}[t]
 \begin{center}
  \begin{tabular}{cccc}
   \hline\hline
   $N$ & 1 & 4 & 12 \\
   \hline\hline
   $\lambda_{hhh}$  &1.32   &1.32     &1.32 \\
   $\lambda_{hhhh}$ &2.9    &2.9      &2.9 \\
   $\lambda_{hss}$  &11.4   &5.02     &2.80 \\
   $\lambda_{hhss}$ &14     &5.6      &3.0 \\
   $\lambda_{ssss}$ &6.5    &1.9      &0.9 \\
   $m_s$[GeV]      &556    &378       &285 \\
   \hline\hline
  \end{tabular}
  \caption{\small The results about the interactions among the Higgs
    boson and singlet scalar bosons derived in
    Ref.~\cite{Endo:2015ifa}.  They are defined in
    Eq.~(\ref{eq:expandVeff}).  Some of them are also input parameters
    of the calculations in Section~\ref{sec:DM}.  These values are
    those in case (I), except for $\lambda_{ssss}$ which is in (II) of
    Table~4 in Ref.~\cite{Endo:2015ifa}.  In our notation of the SM
    Higgs potential the predicted self-couplings are
    $\lambda_{hhh}^\text{(SM)}=\lambda_{hhhh}^\text{(SM)}
    =6\lambda_H=3{m_h}^2/{v_H}^2\simeq0.78$\,.}
 \label{tab:parameter_02}
 \end{center}
\end{table}

\section{Detection of Singlet Scalars}
\label{sec:DM}
\setcounter{equation}{0}

As we discussed the singlet scalars are stable and they can play a
role of dark matter. Since the singlet scalars interact with Higgs
boson, they are thermalized in the early universe if the reheating
temperature is higher than the singlet mass.\footnote{To be specific,
  the reheating temperature should be larger than $300$--$600$~GeV
  (see Table~\ref{tab:parameter_02}), which is a canonical case in the
  early universe.} Then the relic abundance is determined by the
conventional freeze-out scenario. What we need is the annihilation
cross section of $s_i$. Relevant annihilation processes are $s_i s_i
\rightarrow W^+W^-$, $ZZ$, $hh$ and $t\bar{t}$.  The cross sections
for the processes are given by
\begin{align}
  &\sigma_{s_i s_i \rightarrow W^+W^-}=
  \frac{\beta_f(s,{m_W}^2)}{4\pi s\beta_i}
  \left(\frac{\lambda_{hss}\,{m_W}^2}{s-{m_h}^2}\right)^2
  \left[2+\frac{1}{4}\left(\frac{s-2{m_W}^2}{{m_W}^2}\right)^2\right]\ ,
  \\
  &\sigma_{s_i s_i \rightarrow ZZ}=
  \frac{\beta_f(s,{m_Z}^2)}{8\pi s\beta_i}
  \left(\frac{\lambda_{hss}\,{m_Z}^2}{s-{m_h}^2}\right)^2
  \left[2+\frac{1}{4}\left(\frac{s-2{m_Z}^2}{{m_Z}^2}\right)^2\right]\ ,
  \\
  &\sigma_{s_i s_i \rightarrow hh}=
  \frac{1}{16\pi s \beta_i}
  \Biggl[
    {\beta_f}(s,{m_h}^2) \tilde{\lambda}^2 +
    \frac{4\tilde{\lambda}{\lambda_{hss}}^2{v_H}^2}{s\beta_i}\log\frac{t_+}{t_-}
    \nonumber\\ 
    &~~~~~~~~~~~~~~~~~~~
    +
    \frac{2{\lambda_{hss}}^4{v_H}^4}{s\beta_i}
    \left(
    \frac{s\beta_i\beta_f(s,m_h^2)}{t_+t_-} +
    \frac{2}{2{m_h}^2-s}\log\frac{t_+}{t_-}
    \right)
  \Biggr]\ , 
  \\
  &\sigma_{s_i s_i \rightarrow t \bar{t}}=
  \frac{3{\beta_f}^3(s,{m_t}^2)}{8\pi \beta_i}
  \left(\frac{\lambda_{hss}\, m_t}{s-{m_h}^2}\right)^2\ ,
\end{align}
where $m_W$, $m_Z$ and $m_t$ are the masses of $W$, $Z$ and $t$,
respectively. $s$ is the center-of-mass energy in the initial state,
$\beta_i=(1-4{m_s}^2/s)^{1/2}$, $\beta_f(s,m^2)=(1-4m^2/s)^{1/2}$ and
\begin{align}
\tilde{\lambda}
  =\lambda_{hhss} +
  \frac{\lambda_{hhh}\lambda_{hss}{v_H}^2}{s-{m_h}^2}\ ,
  \ \ \ \ \ \ \
t_\pm = {m_h}^2 -\frac{s}{2}\left[1\mp \beta_i \beta_f(s,{m_h}^2)\right]\ .  
\end{align}
 Numerically it is found that the annihilation mode to $t\bar{t}$ is
 subdominant.\footnote{This is expected since the amplitude of fermion
   pair final state is chirality suppressed, {\it i.e.}, the cross
   section is proportional to ${m_t}^2/{m_{s_i}}^4$ instead of
   $1/{m_{s_i}}^2$.} It is straightforward to compute the
 thermal-averaged cross section from above expression. We use the
 formula in Ref.\,\cite{Kolb&Turner} to get the relic abundance. (We
 have checked that the density parameter by solving the Boltzmann
 equation agrees with the approximated result within a few \%.) The
 results are shown in Table~\ref{table:sigma}. (We note that when
 $N\ge 2$ all scalars $s_1$, $\cdots$, $s_N$ become dark matter.)  It
 has figured out that the relic abundance $\Omega_{s_i}$ of $s_i$ is
 much smaller than that of dark matter, which means that $s_i$ cannot
 be the main component of dark matter. This is due to the large
 annihilation cross section enhanced by the large couplings (mainly
 $\lambda_{hss}$). On the other hand, however, $s_i$ can be detected
 in the experiment of direct detection of dark matter, which we will
 discuss below.

For evaluation of the spin-independent cross section of $s_i$ with
nucleon, we adopt the formalism given in Ref.\,\cite{Hisano:2015rsa}.
(See also
Refs.\,\cite{Hisano:2010fy,Hisano:2010ct,Hisano:2011cs,Hisano:2012wm}
for earlier works.) In the present case only scalar-type operators are
induced by Higgs-exchange diagram, then the effective Lagrangian for
the scattering process is
\begin{align}
  {\cal L}_{\rm eff}=\sum_{i=q,G}C^i_{\rm S}{\cal O}^i_{\rm S}\ ,
\end{align}
where
\begin{align}
  {\cal O}^q_{\rm S}=m_q{s_i}^2 \bar{q}q\ , ~~~~~~~~~~~
  {\cal O}^G_{\rm S}=\frac{\alpha_s}{\pi}{s_i}^2G^a_{\mu\nu}G^{a\mu\nu}\ .
\end{align}
$m_q$ is quark mass, $G_{\mu\nu}^a$ is the gluon field strength and
$\alpha_s$ is the strong coupling constant. By integrating out the
Higgs boson (and top quark), the Wilson coefficients at the
electroweak scale $\mu_W \simeq m_Z$ at the next-to-leading order in
$\alpha_s$ are given by
\begin{align}
  &C^q_{\rm S}(\mu_W)= \frac{\lambda_{hss}}{2{m_h}^2}\ ,
  \\
  &C^G_{\rm S}(\mu_W)= -\frac{\lambda_{hss}}{24{m_h}^2}
  \left[1+\frac{11\alpha_s}{4\pi}\right]\ .
\end{align}
The amplitude is given by the hadronic matrix elements, {\it i.e.}
$\langle N |m_q \bar{q}q | N \rangle$, $\langle N |
\frac{\alpha_s}{\pi}{s_i}^2G^a_{\mu\nu}G^{a\mu\nu}| N \rangle$
($N=p,n$), which are obtained from lattice simulations
\cite{Young:2009zb, Oksuzian:2012rzb} and the QCD trace
anomaly~\cite{Shifman:1978zn},\footnote{It is given by
  $\Theta^\mu_{~\mu}=\frac{\beta
    (\alpha_s)}{4\alpha_s}G^a_{\mu\nu}G^{a\mu\nu}_{}
  +(1-\gamma_m^{})\sum_{q}m_q\bar{q}q$, and $m_N=\langle N
  |\Theta^\mu_{~\mu} | N \rangle$ to derive Eq.\,\eqref{eq:G^2}.}
\begin{align}
  \langle N |m_q \bar{q}q | N \rangle &= m_N f_{T_q}^{(N)}\ ,
  \\
  \langle N |
  \frac{\alpha_s}{\pi}G^a_{\mu\nu}G^{a\mu\nu}| N \rangle
  &=m_N\frac{4{\alpha_s}^2}{\pi \beta_s^{\small (N_f=3)}}
  \Bigl[1-(1-\gamma_m)\sum_{q=u,d,s}f_{T_q}^{(N)}\Bigr]\ .
  \label{eq:G^2}
\end{align}
Here $m_N$ is nucleon mass, $f_{T_q}^{(N)}$ is mass fractions, {\it
  e.g.}, $f_{T_u}^{(p)}=0.019(5)$, $f_{T_d}^{(p)}=0.027(6)$ and 
$f_{T_s}^{(p)}=0.009(22)$, which are evaluated in
Ref.\,\cite{Hisano:2012wm} based on Refs.\,\cite{Young:2009zb,
  Oksuzian:2012rzb}. $\beta_s$ and $\gamma_m$ are the beta
function of $\alpha_s$ and the anomalous dimension of quark mass 
defined by $\beta_s=\mu\frac{d\alpha_s}{d\mu}$ and 
$\gamma_m m_q=\mu\frac{dm_q}{d\mu}$, respectively. 
In Eq.\,\eqref{eq:G^2} the number of flavors $N_f=3$ is
taken since we only know the mass fractions for the light quarks,
which means that the Wilson coefficients $C^q_{\rm S}$ and $C^G_{\rm
  S}$ should be evaluated at the hadronic scale $\mu_{\rm had}\simeq
1~{\rm GeV}$. This can be done by the matching procedure at each quark
threshold (bottom and charm quarks) and by solving the renormalization
group equation for the Wilson coefficients. (For the details, such as
the matching and renormalization group evolution at the next-to-leading
order in $\alpha_s$, see Ref.\,\cite{Hisano:2015rsa}). Finally the 
spin-independent cross section is given by
\begin{align}
  \sigma_{\rm SI}^{(N)}=
  \frac{1}{\pi} \frac{{m_N}^2}{(m_s +m_N)^2}|f_{\rm S}^{(N)}|^2\ ,
\end{align}
with
\begin{align}
  f^{(N)}_{\rm S} =\sum_{q=u,d,s}C^q_{\text{S}}(\mu_{\text{had}})
\langle N|m_q\bar{q}q|N\rangle +C^G_{\text{S}}(\mu_{\text{had}})
\langle N|\frac{\alpha_s}{\pi} G^a_{\mu\nu}G^{a\mu\nu}|N\rangle\ .
\end{align}
If renormalization group evolution is ignored as well as taking the
leading order threshold matching, then the effective scattering
amplitude is simply given by
\begin{align}
  \frac{f^{(N)}_{\rm S}}{m_N}|_{\rm est} \approx
  \frac{\lambda_{hss}}{2{m_h}^2}
  \left[\frac{2}{9} +\frac{7}{9} \sum_{q=u,d,s}f_{T_q}^{(N)}\right],
  \label{eq:f_Sest}
\end{align}
which is often used in the literature. To see the impact of the proper
matching procedure to the effective coupling $f_{\rm S}^{(N)}$, we
show the numerical values in Table~\ref{table:f_S}. The difference
between the LO and the NLO results is about $4\%$, while the one
between the rough estimation and the NLO is about $7\%$, which gives
rise to about $14\%$ deviation in the spin-independent cross section.

\begin{table}[t]
 \begin{center}
  \begin{tabular}{cccc}
   \hline \hline
   $N$  & $1$ & $4$ & $12$ \\
   \hline \hline
   $\Omega_{s_i}/\Omega_{\rm DM}$ &
   $2.01 \times 10^{-4}$ & $4.54 \times 10^{-4}$ & $8.07 \times 10^{-4}$ \\
   $\tilde{\sigma}_{\rm SI}^{(p)}$\,[$10^{-46}\,{\rm cm}^{2}$]  &
   6.77 & 25.6 & 74.5 \\
   \hline \hline
  \end{tabular}
  \caption{\small Fractions of the energy density of $s_i$ in the
    observed dark matter density, and spin-independent cross sections
    of the scalars with proton multiplied by their fractions in the
    present dark matter density. The cross sections are computed with 
    the effective couplings $f_\text{S}^{(p)}|_\text{NLO}$ shown in 
    Table \ref{table:f_S}.}
  \label{table:sigma}
 \end{center}
\end{table}

\begin{table}[t]
 \begin{center}
  \begin{tabular}{cccc}
   \hline \hline
   $N$  & $1$ & $4$ & $12$ \\
   \hline \hline
   $f_{\rm S}^{(p)}|_{\rm est}$\,[$10^{-5}\,{\rm GeV}^{-1}$]  &
   9.07 & 3.99 & 2.23 \\
   $f_{\rm S}^{(p)}|_{\rm LO}$\,[$10^{-5}\,{\rm GeV}^{-1}$]   &
   10.2 & 4.49 & 2.50 \\
   $f_{\rm S}^{(p)}|_{\rm NLO}$\,[$10^{-5}\,{\rm GeV}^{-1}$]  &
   9.77 & 4.30 & 2.40 \\
   \hline \hline
  \end{tabular}
  \caption{\small Amplitude for $s_i$-proton scattering.  $f_{\rm
      S}^{(p)}|_{\rm est}$ is given by Eq.\,\eqref{eq:f_Sest} and
    $f_{\rm S}^{(p)}|_{\rm LO}$, $f_{\rm S}^{(p)}|_{\rm NLO}$ are the
    results obtained by appropriate matching at the leading order and
    the next-to-leading order, respectively.}
  \label{table:f_S}
 \end{center}
\end{table}

\begin{figure}[t]
 \begin{center}
  \includegraphics[width=0.7\linewidth]{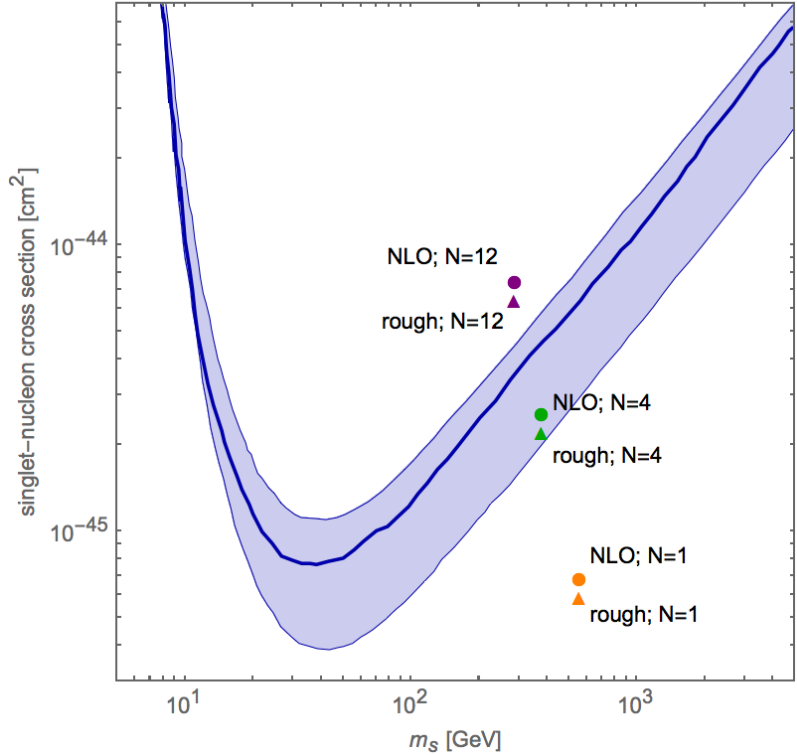}
  \caption{\small Spin-independent cross sections of the singlet
    scalars compared with the LUX 90\%
    C.L. bound~\cite{Akerib:2013tjd}. Orange, green and purple points
    represent $N=1$, 4 and 12 cases, respectively. Triangle and circle
    points represent rough estimation and the next-to-leading order
    calculation, respectively.}
  \label{figure:LUX}
 \end{center}
\end{figure}

Now we are ready to see the experimental consequence of the
model. Since the scalar fields are not the main component of the
present DM, the ``effective'' cross section of the singlet with
nucleon is obtained by multiplying the fraction of the total abundance
of the scalars in the present DM density,
\begin{align}
  \tilde{\sigma}_{\rm SI}^{(N)}= \sigma_{\rm SI}^{(N)}
  \sum_{i=1,\cdots, N}\frac{\Omega_{s_i}}{\Omega_{\rm DM}}\ , 
\end{align}
where $\Omega_{\rm DM}=0.264$~\cite{Ade:2015xua}. The results are
given in Table~\ref{table:sigma} and Fig.~\ref{figure:LUX}. In
Table~\ref{table:sigma} we have used the NLO result.  Compared with
the most stringent bound for the cross section~\cite{Akerib:2013tjd},
$N=12$ case has already been excluded at $90\%$ C.L. The others, {\it
  i.e.} $N=1$ and 4 cases, are still viable, but $N=4$ case is very
close to the present bound. The cross section is far below the present
bound for $N=1$ case. However, it is much larger than the neutrino
background in the direct detection
experiments~\cite{Billard:2013qya}. Therefore, our model (with any
number of $N$) can be tested in ton-scale future experiments, such as
LZ program~\cite{Malling:2011va}. Recall that $N=1$ is favored in
terms of Veltman's condition as well as the fine-tuning of the Higgs
mass, which is discussed in the previous section. Thus the result
shows that the most well-motivated case will be able to be examined in
the future experiments.

\section{Conclusion}
\label{sec:conclusion}
\setcounter{equation}{0} 

In this letter we have studied direct detection of singlet scalar dark
matter in a classically scale-invariant extension of the standard
model.  The model extends the Higgs sector to have an additional
electroweak singlet scalars that form a multiplet of global $O(N)$ 
symmetry, and the electroweak
symmetry is broken via Coleman-Weinberg mechanism. Recently the Higgs
self-couplings as well as new couplings and the singlet mass were
precisely computed in Ref.~\cite{Endo:2015ifa}.  In the work it was
shown the Higgs self-couplings deviate from the standard model
prediction significantly, which can be observed at the next-generation
collider experiments such as the ILC. Another important outcome of
their analysis is unbroken $O(N)$.  Consequently the singlet scalars
are cosmologically stable and can play a role of dark matter.  Since
all couplings and mass parameters are fixed for given number of
$N$~\cite{Endo:2015ifa}, it is possible to precisely predict the
nature of the singlet scalars.  Therefore detection of the singlet
scalars is complementary for the test of the model.

Assuming that the reheating temperature is above the singlet mass, we
have computed the thermal relic abundance of the singlet scalars,
and the scattering amplitude of the scalars with nucleon. For the
precise determination of the scattering cross section we have used the
formalism~\cite{Hisano:2015rsa} which takes into account the
next-to-leading order QCD effect in consistent way. We have focused on
three benchmarks, {\it i.e.} $N=1$, $4$ and $12$ (the singlet masses
are predicted as $556$, $378$ and $285$ GeV, respectively). Then it
has been found that although the relic abundance is much smaller than
the present dark matter ($\Omega_{s_i}/\Omega_\text{DM}\sim
\mathcal{O}(10^{-4})$), the scattering rate is enhanced due to the
large Higgs-singlet coupling. To be concrete, $N=12$ case has already 
been excluded by the LUX experiments, meanwhile $N=4$ case is near the
bound. In $N=1$ case which is favored in terms of the fine-tuning
regarding the Higgs mass, the effective spin-independent cross section
($\simeq 6.8\times10^{-46}\,\text{cm}^2$) is far below the current
bound.  It is, however, much larger than the neutrino background. Thus
it is concluded that the whole parameter space of this scenario is
testable in the future ton-scale detector of dark matter direct
detection.

\section*{Acknowledgement}

The authors would like to express their special thanks to Yukinari
Sumino for many helpful comments and careful reading of this paper.
This work was supported in part by the German Science Foundation (DFG) 
within the Collaborative Research Center 676 ``Particles, Strings 
and the Early Universe'' (K.I.).

\appendix 
\section{Parameters in the effective potential}
\label{sec:pot_param}

Here is the list of parameters of the one-loop
potential in Eq.~\eqref{eq:effective_pot_unimp_complete}:
\begin{align}
&n_W=6\,,~~
{M_W}^2
=\frac{1}{4}g^2{\phi}^{2}\,,~~
c_W=\frac{5}{6}\,;
\nonumber\\
&n_Z=3\,,~~
{M_Z}^2
=\frac{1}{4}(g^2+{g'}^2){\phi}^{2}\,,~~
c_Z=\frac{5}{6}\,;
\nonumber\\
&n_t=-12\,,~~
{M_t}^2
=\frac{1}{2}{y_t}^2
{\phi}^{2}\,,~~
c_t=\frac{3}{2}\, ;
\nonumber\\
&n_\pm=1\,,~~
{M_\pm}^2
=F_\pm\,,~~
c_\pm=\frac{3}{2}\,;
\nonumber\\
&n_{H_\text{NG}}=3\,,~~
{M_{H_\text{NG}}}^2
=\lambda_{H}{\phi}^2+\lambda_{HS}{\varphi}^2\,,~~
c_{H_\text{NG}}=\frac{3}{2}\,;
\nonumber\\
&n_{S_\text{NG}}=N-1\,,~~
{M_{S_\text{NG}}}^2
=\lambda_{HS}{\phi}^2+\lambda_{S}{\varphi}^2\,,~~
c_{S_\text{NG}}=\frac{3}{2}\, ,
	\label{internal-modes}
\end{align}
where $i=W,\,Z,\,t$ show $W,\,Z,\,t$ in the loop, respectively.
$i=\pm$ indicates $\varphi$, $\phi$, while $i=H_{\rm NG},\,S_{\rm NG}$
stand for the degrees of freedom which are orthogonal to $\phi$ and
$\varphi$, respectively.  In the effective potential with the precise
order counting ({\it i.e.}
Eq.\,\eqref{eq:0th+1st_perturbation_order}), we use
\begin{align}
F_{\pm\text{app}}(\phi,\,\varphi) =&
\frac{\lambda_{HS}}{2}{\phi}^2
+\frac{\lambda_{HS}+3\lambda_{S}}{2}{\varphi}^2
\nonumber\\
&\pm\sqrt{\left[-\frac{\lambda_{HS}}{2}{\phi}^2
    +\frac{\lambda_{HS}-3\lambda_{S}}{2}{\varphi}^2
    \right]^2+4{\lambda_{HS}}^2{\phi}^2{\varphi}^2} \, ,
\label{eq:Fapp}
\end{align}
for $F_{\pm}$.  The reason for dropping $F_{-\text{app}}$ in
Eq.\,\eqref{eq:0th+1st_perturbation_order} is explained in
Ref.\,\cite{Endo:2015ifa} (see also Ref.\,\cite{Martin:2014bca}).

\providecommand{\href}[2]{#2}\begingroup\raggedright\endgroup
 
\end{document}